\documentclass[aps, showkeys,showpacs,nofootinbib,floatfix]{revtex4}

\usepackage{amssymb}
\usepackage{amsfonts}
\usepackage{amsmath}
\usepackage{graphicx}
\usepackage{hyperref}
\usepackage{graphicx}

\begin{document}

\title{Logarithmic Entropy of Kehagias-Sfetsos black hole with Self-gravitation in Asymptotically Flat IR Modified Ho$\check{r}$ava Gravity}
\author{Molin Liu}
\thanks{Corresponding author\\E-mail address: mlliu@mail2.xytc.edu.cn}
\author{Junwang Lu}
\affiliation{College of Physics and Electronic Engineering,
Xinyang Normal University, Xinyang, 464000, P. R. China}

\begin{abstract}
Motivated by recent logarithmic entropy of Ho$\check{r}$ava-Lifshitz gravity, we investigate Hawking radiation for Kehagias-Sfetsos black hole from tunneling perspective.
After considering the effect of self-gravitation, we calculate the emission rate and entropy of quantum tunneling by using Kraus-Parikh-Wilczek method. Meanwhile, both massless and massive particles are considered in this letter. Interestingly, two types tunneling particles have the same emission rate $\Gamma$ and entropy $S_b$ whose analytical formulae are $\Gamma = \exp{\left[\pi \left(r_{in}^2 - r_{out}^2\right)/2 + \pi/\alpha \ln r_{in}/r_{out}\right]}$ and $S_b = A/4 + \pi/\alpha \ln (A/4)$, respectively. Here, $\alpha$ is the Ho$\check{r}$ava-Lifshitz field parameter.
The results show that the logarithmic entropy of Ho$\check{r}$ava-Lifshitz gravity could be explained well by the self-gravitation, which is totally different from other methods. The study of this semiclassical tunneling process may shed light on the understand of Ho$\check{r}$ava-Lifshitz gravity.
\end{abstract}



\maketitle

\section{Introduction}
Hawking radiation is one of the most important predictions in black hole physics. In the early researches, it was looked as the pure emission thermal spectrum which leads to the information loss and the breakdown of unitary theory. The main reasons are relevant two points: one is the potential of tunneling is not found; another is the reaction of particle is not considered.
Subsequently, one semiclassical model was presented in Refs.\cite{Keski1,Kraus1,Kraus2,Keski2} in which Hawking radiation is considered as a tunneling process across the horizon for the static spherically symmetric (SSS) black hole. Recently, Parikh and Wilcze have put forward it by considering the self-gravitation of particles \cite{ParikhWilczek}. Such approaches are called Kraus-Parikh-Wilczek (KPW) methodology in the literature. The emission spectrum of black hole radiance is not pure thermal and the effect of self-gravitation gives rise to semiclassical correction to the black hole entropy. The actual emission spectrum is related with the change of entropy during radiating particle. Hence, KPW method satisfies unitary theory of quantum mechanics and supports the conversation of information. There are many works focused on the application of tunneling process in various systems \cite{yingyong1old,yingyong2new}.

On the other hand, Ho$\breve{r}$ava recent proposes a power-counting renormalizable, ultraviolet (UV) complete theory of gravity at the Lifshitz point \cite{Horava}. It may be regarded as a UV complete candidate for general relativity. The basic property of such a theory
is the invariance under the anisotropic rescaling, $t \longrightarrow l^z t$, $\overrightarrow{x}\longrightarrow l \overrightarrow{x}$. By the deformed
rescaling with an appropriate $z$, the action turns out to be power counting
renormalizable. Since then, more attentions have been paid to its black hole solutions \cite{Kehagias,Lu,Cai11,cai22,Colgain,Cai33,Myung2,park,Myung1}. Kehagias-Sfetsos (KS) have obtained the ``$\lambda=1$" black hole solution
in asymptotically flat spacetimes \cite{Kehagias}. Lu-Mei-Pope (LMP) have
found (A)dS Schwarzschild black hole with with dynamical parameter $\lambda$ \cite{Lu}. Cai-Cao-Oha (CCO) have gotten topological (charged) black holes with an arbitrary constant
scalar curvature horizon \cite{Cai11}. Park has obtained the black hole for arbitrary cosmological constant\cite{park}, and so on.

Since the KS black hole is presented \cite{Kehagias}, its thermodynamics attracts much attention. Among them, the black hole entropy is obtained by the first law of thermodynamics, which has a logarithmic form as follows,
\begin{equation}\label{ksciteentropy}
    S = \frac{A}{4} + \frac{\pi}{\omega}\ln (\frac{A}{4}),
\end{equation}
which is completely different from the Bekenstein-Hawking area entropy \cite{Myung1}.
About its explanation, there are two versions. One is Myung's generalized uncertainty principle (GUP) inspired model \cite{YSMyung22}, and the
other is Cai and Ohta's casting model \cite{Caic111}. In Myung's
model \cite{YSMyung22}, it is
considered as the GUP quantum correction entropy. In Cai and Ohta's casting model \cite{Caic111}, the gravitational field equation can be cast to a form of the first law of thermodynamics at the black hole horizon, in which logarithmic entropy could be looked as the casting result.

Considering all above factors, we wonder what happens if the self-gravitation effect is involved for HL gravity. Especially, whether it can give the logarithmic entropy in the tunneling picture. So, motivated by the situations above, we apply the KPW method \cite{ParikhWilczek} to KS black hole in HL gravity. The total (ADM) mass keep fixed, while the mass of the KS black hole decreases due to the emitted radiation.

This letter is organized as follows. In section II, we present the Kehagias-Sfetsos black hole in brief. In section III, we calculate the Hawking radiation emission rate and the black hole entropy for the massless particles. In section IV, we recalculate the emission rate and the black hole entropy for the massive particles. Section V is the conclusions. We adopt the signature $(-, +, +, +)$ and put $\hbar$, $c$, and $G$ equal to unity.
\section{asymptotically flat black hole in deformed Ho$\check{r}$ava-Lifshitz gravity}
In this section, we review briefly the KS black hole for HL gravity. Firstly, this type solution is under
the limit of $\Lambda_{W} \longrightarrow 0$ with running constant
$\lambda = 1$ in the IR critical point $z = 1$. The spacetime geometric
is parameterized with ADM formalism,
\begin{equation}\label{ADMmetric}
    d s^2 = - N^2 d t_{L}^2 + g_{ij} \left( d  x^i + N^i d t_{L}\right)\left(d x^j + N^j dt_{L}\right).
\end{equation}
The action for the fields of HL theory is

\begin{eqnarray}\label{action1}
\nonumber S &=& \int dt_{L} d^3 x \sqrt{g} N \bigg{\{} \frac{2}{\kappa^2} \left(K_{ij}K^{ij} - \lambda K^2 \right) - \frac{\kappa^2}{2 \alpha^4} C_{ij} C^{ij} + \frac{\kappa^2 \mu}{2 \alpha^2} \epsilon^{ijk} R_{il}^{(3)} \nabla_{j} R^{(3)l}_{\ \ \ \ k}\\
    && \ \ \ \ \ \ \ \ \ \ \ \ \ \ \ \ \ \ \ -\frac{\kappa^2 \mu^2}{8} R_{ij}^{(3)}R^{(3)ij} + \frac{\kappa^2 \mu^2}{8 \left(1 - 3\lambda\right)} \left(\frac{1 - 4 \lambda}{4}\left(R^{(3)}\right)^2 + \Lambda_{W} R^{(3)} - 3 \Lambda^2_{W}\right) + \mu^4 R^{(3)}\bigg{\}}.
\end{eqnarray}
The second fundamental form of extrinsic curvature $K_{ij}$, and
the Cotton tensor $C^{ij}$ are given as follows,
\begin{eqnarray}
  K_{ij} &=& \frac{1}{2N} \left(\frac{\partial }{\partial t_{L}}g_{ij} - \nabla_i N_j - \nabla_j N^i\right), \label{extrinsiccurvature}\\
  C^{ij} &=& \epsilon^{ikl} \nabla_k \left(R^{(3)j}_{l} - \frac{1}{4} R^{(3)} \delta_{l}^{j}\right), \label{Cottontensor}
\end{eqnarray}
where $\kappa$, $\lambda$, $\alpha$ are dimensionless coupling constants and $\mu$, $\Lambda_{W}$ have the mass dimensions $[\mu] = 1$, $[\Lambda_{\alpha}] = 2$. The last term of metric (\ref{action1}) represents a soft violation of the detailed balance condition. In the limit of $\Lambda_{W} \longrightarrow 0$, we can obtain a deformed action as,
\begin{equation}
    S = \int dt_{L} d^3 x \left(\mathcal{L}_0 + \mathcal{L}_1\right),\label{deformedaction1}
\end{equation}
where
\begin{eqnarray}
    \mathcal{L}_0 &=& \sqrt{g} N \bigg{\{}\frac{2}{\kappa^2} \left(K_{ij}K^{ij} - \lambda K^2\right)\bigg{\}},\label{deformedaction2}\\
    \mathcal{L}_1 &=& \sqrt{g} N \bigg{\{} \frac{\kappa^2 \mu^2 \left(1 - 4\lambda\right)}{32\left(1 - 3\lambda\right)} \mathcal{R}^2 - \frac{\kappa^2}{2 \alpha^4}\left(C_{ij} -\frac{\mu \alpha^2}{2} R_{ij}\right) \left(C^{ij}-\frac{\mu \alpha^2}{2} R^{ij}\right) + \mu^4 \mathcal{R} \bigg{\}}.\label{deformedaction3}
\end{eqnarray}
Comparing above parameters with the usual GR ADM formalism, we can obtain the speed of light $c$, the Newton's constant $G$ and the cosmological constant $\Lambda$, which are listed as,
\begin{equation}\label{cGlambda}
    c = \frac{\kappa^2 \mu}{4} \sqrt{\frac{\Lambda_{W}}{1 - 3\lambda}},\ \ G = \frac{\kappa^2}{32\pi c},\ \ \Lambda = \frac{3}{2} \Lambda_{W}.
\end{equation}
For $\lambda = 1$ case with $\alpha = 16 \mu^2/\kappa^2$, an asymptotically flat black hole solution is presented by Kehagias and Sfetsos \cite{Kehagias} which could be treated as the counterpart of Schwarzschild black hole of GR. The spherically symmetric metric of KS black hole is given as follows,
\begin{equation}\label{Kehagiassolution}
    d s^2 = -f(r) d t_{L}^2 + \frac{d r^2}{f(r)} + r^2 \left(d \theta^2 + \sin^2 \theta d \varphi^2 \right).
\end{equation}
The lapse function is
\begin{equation}\label{metricfunction}
    f(r) = 1 + \alpha r^2 - \sqrt{r\left(\alpha^2 r^3 +4 \alpha M\right)},
\end{equation}
where the parameter $M$ is an integration constant related with black hole mass.

Using the null hypersurface condition, we can find two horizons in this spacetime, which are the inner $r_-$ and the outer event horizon $r_+$,
\begin{equation}\label{horizons}
    r_{\pm} = M \left(1 \pm \sqrt{1 - \frac{1}{2 \alpha M^2}}\right).
\end{equation}
Recently, the parameter $M$ could be assumed as the ADM mass $M_{KS}$ by Myung in Ref \cite{Myung1}. Then, we can get the Hawking temperature $T_{KS}$ and
heat capacity $C_{KS}$ of KS BH, which are listed as follows,
\begin{eqnarray}
  M_{KS} &=& \frac{1 + 2\alpha r\pm^2}{4\alpha r\pm}, \label{ADSmass}\\
  T_{KS} &=& \frac{2\alpha r_+^2 - 1}{8\pi r_+(\alpha r_+^2 + 1)}, \label{temperature}\\
  C_{KS} &=& -\frac{2\pi}{\alpha} \left[\frac{(\alpha r_+^2 + 1)^2 (2\alpha r_+^2 - 1)}{2\alpha^2 r_+^4 - 5\alpha r_+^2 - 1}\right].\label{heatcapacity}
\end{eqnarray}
According to the first thermodynamics law $d M_{KS} = T_{KS} d S_{KS}$, the entropy is derived as
\begin{equation}\label{ksentropy}
    S_{KS} = \frac{A}{4} + \frac{\pi}{\alpha}\ln \left(\frac{A}{4}\right).
\end{equation}
The last logarithmic term of r.h.s. could be treated as the correction of generalized uncertainty principle \cite{YSMyung22}. Meanwhile, it could also be looked as one result of casting the gravitational field equation to the first law of thermodynamics at horizon \cite{Caic111}.
\section{logarithmic entropy from quantum tunneling of massless particles}
In this section, we will calculate the emission spectrum and the entropy for the massless particles in the tunneling picture. Before performing computation, we need assume that the KS solution still retains the general coordinate invariance. It is supported by many literatures, for instance the tortoise coordinate transformation of quasinormal modes \cite{1qnm1,1qnm2,1qnm3} and so on. Then, we can present the Painlev$\acute{e}$ type coordinate transformation of KS BH. Because the KS BH has a characteristic of SSS geometry, the infinity red shift surface is coincident with the event horizon. So it supports the geometrical optical approximation. In order to eliminate the coordinate singularity at the event horizon $r_+$, we adopt the Painlev$\acute{e}$ type coordinate conversion \cite{Painleve},
\begin{equation}\label{painleve}
    d t = d t_{L} + \frac{\sqrt{1 - f(r)}}{f(r)} d r,
\end{equation}
where $t_{L}$ is a Schwarzschild type time appeared in original line element (\ref{Kehagiassolution}). The new form of Painlev$\acute{e}$ time $t$ above could help us well describe the tunneling process. Submitting Eq.(\ref{painleve}) into metric (\ref{Kehagiassolution}), we can get a new element line without coordinate singularity as,
\begin{equation}\label{pbianhuanmetric}
d s^2 = - \left[1 + \alpha r^2 - \sqrt{r\left(\alpha^2 r^3 +4 \alpha M\right)}\right] d t^2 + 2 \sqrt{\sqrt{r\left(\alpha^2 r^3 +4 \alpha M\right)} - \alpha r^2} d t d r + d r^2 + r^2 d \Omega^2.
\end{equation}
In this spacetime (\ref{pbianhuanmetric}), the energy conservation is still in effect and the time-like Killing vector also exists. Each hypersurface equal in time could be treated as the flat Euclidean spacetime. The observer at infinity does not distinguish whether it is static or not. Obviously, this spacetime (\ref{pbianhuanmetric}) is reduced to the static case at infinity. Hence, the tunneling process can be utilized more efficiently with the help of above metric (\ref{pbianhuanmetric}).

Then, we turn toward the massless particle radiation. By using the null conditions $ds^2 = 0 = d \Omega^2$, its radial geodesic is written as,
\begin{equation}\label{2.11}
\dot{r} = \frac{d r}{d t} = \pm 1 - \sqrt{1 - f(r)},
\end{equation}
where the signs $+$ and $-$ are corresponding to the outgoing and ingoing geodesics, respectively. Then, the total ADM mass is fixed and only the mass $M$ of the KS black hole changes. When the particle with energy $\omega$ (or mass) radiates outwards the horizon $r_+$, the KS black hole mass is reduced to $M - \omega$. Then, the massless particles travel on a modified geodesics as,
\begin{equation}\label{2.12mgeo}
\dot{r} =  1 - \sqrt{\sqrt{r(\alpha^2 r^3 + 4 \alpha (M - \omega))} - \alpha r^2}.
\end{equation}

According to the simple relation of the emission rate $\Gamma$ and the action of particles $S$,
\begin{equation}\label{213emissionrate}
\Gamma = e^{-2 Im S},
\end{equation}
we can know that the imaginary part of the action is the key point to get tunneling probability. When the positive-energy particle
crosses the event horizon $r_+$, the appropriate radial
coordinate changes in the range of $r \in [r_{out}, r_{in}]$,
\begin{eqnarray}
  r_{in} &=& M \left(1 + \sqrt{1 - \frac{1}{2M^2\alpha}}\right),\label{rin} \\
  r_{out} &=& (M-\omega) \left(1 + \sqrt{1 - \frac{1}{2(M-\omega)^2\alpha}}\right).\label{rou}
\end{eqnarray}
Hence, the imaginary part of the action is
\begin{equation}\label{211imaginary}
    Im S = Im \int_{r_{in}}^{r_{out}} p_r d r = Im \int_{r_{in}}^{r_{out}} \int_{0}^{p} d p dr.
\end{equation}
In order to convert the integral from momentum to energy, the Hamilton equation must be used as
\begin{equation}\label{211hamilton}
    \dot{r} = \frac{d H}{d p}.
\end{equation}
Hence, the momentum variable $d p$ is transmitted to the energy variable $d H$. By using the modified geodesics equation (\ref{2.12mgeo}) and Hamilton equation (\ref{211hamilton}), the integral (\ref{211imaginary}) is reduced to a solvable formula as,
\begin{equation}\label{211image}
   Im S =  Im \int_{r_{in}}^{r_{out}} \int_{M}^{M-\omega} \frac{d (M - \omega)}{1 - \sqrt{\sqrt{r(\alpha^2 r^3 + 4 \alpha (M - \omega))} - \alpha r^2}} d r.
\end{equation}
Then, we adopt a new variable $u$ as follows,
\begin{equation}\label{211nnewu}
    \sqrt{\sqrt{r(\alpha^2 r^3 + 4 \alpha (M - \omega))} - \alpha r^2} = u,
\end{equation}
which leads a simple new relationship about particles energy as,
\begin{equation}\label{211nnedenergy}
  d (M - \omega) = \frac{1}{\alpha r} (u^3 + \alpha r^2 u) d u.
\end{equation}
Submitting Eqs.(\ref{211nnewu}) and (\ref{211nnedenergy}) into Eq.(\ref{211image}), the
imaginary part of action is rewritten as,
\begin{eqnarray}\label{212imapart}
 \nonumber   Im S &=& Im \int_{r_{in}}^{r_{out}} \frac{1}{\alpha r} \int_{u_{in}}^{u_{out}} \frac{u^3 + \alpha r^2 u}{1 - u} d u d r\\
     &=& - Im \int_{r_{in}}^{r_{out}} \frac{1}{\alpha r} \int_{u_{in}}^{u_{out}} \left[\mathcal{F}_{0}(u) + \mathcal{F}_1 (u)\right] du dr,
\end{eqnarray}
where the integrands $\mathcal{F}_{0}(u)$ and $\mathcal{F}_1(u)$ are given as,
\begin{eqnarray}
  \mathcal{F}_{0}(u) &=& u^2 + u + \alpha r^2 + 1, \\
  \mathcal{F}_1(u) &=& \frac{\alpha r^2 + 1}{u - 1}.
\end{eqnarray}
Obviously, the former $\mathcal{F}_{0}(u)$ has no contribution to the imaginary part of action when we take into account $u$ in the range of $[u_{in}, u_{out}]$. So only the last $\mathcal{F}_1 (u)$ works to calculate $Im S$ (\ref{212imapart}). It should be noticed that there is a singular point at $u = 1$ in integrand $\mathcal{F}_1(u)$, which is corresponding to the event horizon position $r = r_+$. The contour integration is adopted on the upper half of complex plane $u$. By using the replacement $u - 1 = \rho e^{i x}$ with $x \in [0, \pi]$, the imaginary part reduces to a simple formal as,
\begin{eqnarray}\label{211impart}
   \nonumber Im S &=& - Im \int_{r_{in}}^{r_{out}}\frac{\alpha r^2 + 1}{\alpha r} \int_{0}^{\pi} \frac{1}{\rho e^{i x}} du dr\\
   \nonumber &=& - \frac{\pi}{\alpha} Im\int_{r_{in}}^{r_{out}}  i \frac{\alpha r^2 + 1}{r} dr\\
  &=& -\pi \left(\frac{r^2}{2} + \frac{1}{\alpha} \ln r\right)\bigg|_{r_{in}}^{r_{out}} = \pi \left(\frac{r_{in}^2 - r_{out}^2}{2} + \frac{1}{\alpha} \ln \frac{r_{in}}{r_{out}}\right).
\end{eqnarray}
Then, according to Eq.(\ref{213emissionrate}), we can obtain the emission rate for the outgoing positive-energy particles,
\begin{equation}\label{2111emissionrate}
    \Gamma = \exp{\left[-\pi \left(\frac{r_{in}^2 - r_{out}^2}{2} + \frac{1}{\alpha} \ln \frac{r_{in}}{r_{out}}\right)\right]}.
\end{equation}

It is well known that the emission rate shown below can be expressed by the temperature and the entropy of source,
\begin{equation}\label{211emissionrate}
    \Gamma = e^{-\beta \omega} = e^{\Delta S_b}.
\end{equation}
Comparing Eqs.(\ref{2111emissionrate}) and (\ref{211emissionrate}), we get the change of entropy as,
\begin{eqnarray}
\nonumber    \Delta S_b &=& S_b (\omega, M, \alpha) - S_b (M, \alpha)\\
 &=&  \left(\pi r_{out}^2 - \pi r_{in}^2\right) + \frac{\pi}{\alpha} \left(\ln \pi r_{out}^2 - \ln \pi r_{in}^2\right).\label{21212changentropy}
\end{eqnarray}
Due to the specific modelling of the self-gravitation, the modified entropy of KS black hole is reduced to the final logarithmic entropy as,
\begin{equation}\label{21213entropy}
    S_b = \frac{A}{4} + \frac{\pi}{\alpha} \ln \frac{A}{4},
\end{equation}
which is identical with the entropy obtained via the first law of thermodynamics in Ref. \cite{Myung1}. Though we obtain a satisfied result, it is confined to massless case only. In the next section, we will analyse the tunneling of massive particles.
\section{logarithmic entropy from quantum tunneling of massive particles}
In this section, we will reexamine the quantum tunneling for massive particles case.
As we know that, the geodesic of massive particle is time-like. So, Eq.(\ref{2.11}) is no longer functions to time-like case more. If we treat the radiative particles near horizon as the spherical de Broglie wave (s wave) \cite{zhang1}, the time-like geodesic could be given as,
\begin{equation}\label{301cedixian}
    \dot{r} = v_p = \frac{v_g}{2} = -\frac{g_{tt}}{2g_{tr}},
\end{equation}
where $v_g$ is the  group velocity and $v_p$ is the phase velocity given by following formulas,
\begin{eqnarray}
  v_g &=& \frac{d r_c}{d t} = \frac{d \omega}{d k}, \\
  v_p &=& \dot{r} = \omega/k.
\end{eqnarray}

Submitting metric (\ref{pbianhuanmetric}) into above Eq.(\ref{301cedixian}), the time-like geodesic for massive particles reduces to a new form as,
\begin{equation}\label{302leishicedixian}
    \dot{r} = \frac{f(r)}{\sqrt{1 - f(r)}} = \frac{1 + \alpha r^2 - \sqrt{r\left(\alpha^2 r^3 + 4 M \alpha\right)}}{2\sqrt{\sqrt{r^2 \left(\alpha^2 r^3 + 4 M \alpha\right)}-\alpha r^2}}.
\end{equation}
Hence, the imaginary part of the action for radiating massive particles is given as,
\begin{equation}\label{303xubu}
    Im S = Im \int_{r_{in}}^{r_{out}} \int_{M}^{M - \omega} \frac{d H}{\dot{r}} dr.
\end{equation}

Submitting Eq.(\ref{302leishicedixian}) into Eq.(\ref{303xubu}), we rewrite the imaginary part $Im\ S$ in a solvable formal as,
\begin{eqnarray}\label{304xxbubu}
  \nonumber Im \ S &=& Im \int_{r_{in}}^{r_{out}} \int_{u_{in}}^{u_{out}}  \frac{-2}{r \alpha} \cdot \frac{u^4 + \alpha r^2 u^2}{u^2 - 1} du dr\\
  &=& Im \int_{r_{in}}^{r_{out}} \frac{-2}{r \alpha}\int_{u_{in}}^{u_{out}} \left[\mathcal{F}_0'(u) + \mathcal{F}_0''(u) + \mathcal{F}_2(u)\right] d u d r,
\end{eqnarray}
where the replacement (\ref{211nnewu}) is adopted. The integrands $\mathcal{F}_0'(u)$, $\mathcal{F}_0''(u)$ and $\mathcal{F}_2(u)$ are listed as,
\begin{eqnarray}
  \mathcal{F}_0'(u) &=& u^2 + \alpha r^2 + 1, \label{30f0pie}\\
   \mathcal{F}_0''(u) &=&  - \frac{\alpha r^2 + 1}{2(u + 1)}, \label{30f02pie}\\
  \mathcal{F}_2(u) &=& \frac{\alpha r^2 + 1}{2(u - 1)}.\label{30f2}
\end{eqnarray}

Here, the continuity of $\mathcal{F}_0'(u)$, $\mathcal{F}_0''(u)$ and
$\mathcal{F}_2(u)$ need to be clarified. The first $\mathcal{F}_0'(u)$ (\ref{30f0pie})
is analytic in all range of variable $u$. The second $\mathcal{F}_0''(u)$ (\ref{30f02pie}) is still analytic, even
though there is a singular point at $u = -1$ but this point is out of the
range of $u \in [u_{out}, u_{in}]$. Hence, the corresponding
integrands $\mathcal{F}_0'(u)$ and $\mathcal{F}_0''(u)$ about $u$
have no contribution to the imaginary part of particles action.
Finally, only the third part $\mathcal{F}_2(u)$ is effective to the imaginary part.

Submitting $\mathcal{F}_2(u)$ (\ref{30f2}) into Eq.(\ref{304xxbubu}),
we can find $Im S$ could reduce to a simple form as,
\begin{equation}\label{30xubujidenaa}
    Im S = - Im \int_{r_{in}}^{r_{out}} \frac{\alpha r^2 + 1}{r
    \alpha} \int_{u_{in}}^{u_{out}} \frac{1}{u - 1} du dr.
\end{equation}
It is interesting that above $Im S$ (\ref{30xubujidenaa}) is accordance with the result of massless case, namely Eq.(\ref{212imapart}). Then, left in this section, we can adopt the similar process route as former massless case. With the help of the contour integration about variable $u$, the final result of $Im S $ for massless particles are obtained as
\begin{equation}\label{massiveimaction}
    Im S = \pi \left(\frac{r_{in}^2 - r_{out}^2}{2} + \frac{1}{\alpha} \ln \frac{r_{in}}{r_{out}}\right).
\end{equation}
So, the emission rate of massive particle is
\begin{equation}\label{massiveemissionrate}
\Gamma = \exp{\left[-\pi \left(\frac{r_{in}^2 - r_{out}^2}{2} + \frac{1}{\alpha} \ln \frac{r_{in}}{r_{out}}\right)\right]},
\end{equation}
and the entropy is
\begin{equation}\label{3030entropy}
    S_b = \frac{A}{4} + \frac{\pi}{\alpha} \ln \frac{A}{4}.
\end{equation}
Hence, the tunneling particles have the same emission rate and entropy whatever they are massless or massive. Most importantly, the entropies (\ref{21213entropy}) and (\ref{3030entropy}) are indeed the logarithmic entropy when the self-action is considered.

\section{conclusion}
In this letter, the semiclassical tunneling process of KS black hole is investigated by the KPW methodology. We have calculated the quantum tunneling rate and entropy for massless and massive particles, respectively. It is found that the logarithmic entropy is explained well by self-gravitation in Hawking radiation. Now, we summarize what has been achieved.

1. For the massless particle radiation, the radial null geodesic is obtained through the Painlev$\acute{e}$ type line element (\ref{pbianhuanmetric}) without coordinate singularity. If we fix the ADM mass of spacetime and treat the particles radiated as a $s$ wave with energy $\omega$, the mass parameter in Eqs.(\ref{pbianhuanmetric}) and (\ref{2.11}) should be changed from $M$ to $M-\omega$ based on the self-gravitation action. Naturally, the imaginary part of action is obtained through a contour integral about variable $u$. It should be noticed that $\mathcal{F}_{0}(u)$, which is analytic in the range of $[u_{in}, u_{out}]$, has no contribution to the imaginary part of massless particles' action. Then we can get the emission rate and the entropy change for massless particle radiation.

2. For the massive particle radiation, the Eq.(\ref{2.11}) is no longer applicable due to $d s \neq 0$ for massless particles. We must find another way to perform our tunneling analysis unlike the massless case. In order to find the time-like geodesic equation we adopt Zhang and Zhao's formulae relevant group velocity and phase velocity \cite{zhang1}. By the contour integral of action imaginary part, we can get the imaginary part of action for massless particles. It is found that the effective part of the imaginary part Eq.(\ref{30xubujidenaa}) is accordance with that of massless case, namely Eq.(\ref{212imapart}). Hence, whether the particles are massless or massive, the emission spectrum and entropy have the same formulae in the tunneling process.

3. Overall considering the tunneling analysis about two types particles, we may find something as follows. The background spacetime should have quantum vacuum fluctuations when black hole is radiating particles. This effect is more significant for HL gravity being a candidature of quantum gravity. The actual system investigated should include the HL gravity field and radiated particles. In another words, when we refer to the thermodynamics of HL black hole, both radiation and gravity degrees should be included in all. The HL gravity field should be treated as dynamical one. If we consider the reaction of particles radiated on background, the actual emission spectrum should deviate from pure thermal case. The particle of Hawking radiation takes out part information of black hole. The emission spectrum is relevant to the entropy change therefore. The entropy obtained satisfies the unitary principle of quantum mechanics. Meanwhile, it also supports the conservation of information. At the end, KS black hole indeed has the logarithmic entropy, if the radiation is treated as the semiclassical quantum tunneling process.
\acknowledgments  Project is supported by the Natural Science
Foundation of P.R. China (No.11005088).

\end{document}